\documentclass[onecolumn,prl]{revtex4}
\usepackage{graphicx}
\usepackage{subfigure}
\usepackage{amssymb,amsmath}
\usepackage[usenames]{color}

\def\0{{(0)}}
\def\1{{(1)}}

\def\n{\textbf{n}}
\def\N{\textbf{N}}
\def\x{\textbf{x}}

\def\qsm{q_{\textrm{sm}}}
\def\q{q_0}
\def\sn{\textrm{sn}}
\def\cn{\textrm{cn}}
\def\dn{\textrm{dn}}

\def\am{\textrm{am}}

\def\Re{\textrm{Re}}
\def\Im{\textrm{Im}}

\def\ex{{\bf e}_x}
\def\ey{{\bf e}_y}
\def\ez{{\bf e}_z}

\def\sgn{\textrm{sgn}}
\def\sq{{\rm sq}}
\def\hex{{\rm hex}}

\begin{document}

\title{Straight Round the Twist: Frustration and Chirality in Smectics-A}

\author{Elisabetta A. Matsumoto}
\affiliation{School of Physics, Georgia Institute of Technology, 837 State Street, Atlanta GA 30309, USA}
\email{sabetta@gatech.edu}
\author{Randall D. Kamien}
\affiliation{Department of Physics and Astronomy, University of Pennsylvania, 209 S. 33rd St., Philadelphia, Pennsylvania, USA}
\author{Gareth P. Alexander}
\affiliation{Department of Physics and Centre for Complexity Science, University of Warwick, Coventry CV4 7AL, UK}

\date{\today}

\begin{abstract}
Frustration is a powerful mechanism in condensed matter systems, driving both order and complexity.  In smectics, the frustration between macroscopic chirality and equally spaced layers generates textures characterised by a proliferation of defects.  In this article, we study several different ground states of the chiral Landau-de Gennes free energy for a smectic liquid crystal.  The standard theory finds the twist grain boundary (TGB) phase to be the ground state for chiral type II smectics.  However, for very highly chiral systems, the hierarchical helical nanofilament (HN) phase can form and is stable over the TGB.  

\bigskip

\bigskip

\end{abstract}

\maketitle

\section{Introduction}

Topological defects often characterise a particular phase of a condensed matter system as they are defined by the interface between regions with different symmetries.  Although typically the hallmark of a phase transition, topological defects can also arise from frustration within a system.  They act to mediate the frustration induced by two competing yet mutually exclusive terms in the free energy.  The Abrikosov phase of type II superconductors might be the most familiar example of this.  A lattice of flux vortices reconciles the Meissner effect of the superconducting state with an applied magnetic field. Such frustration exists in soft matter systems and frequently results in complex geometrical states of matter. The prototypical example is the chiral smectic-A${^*}$, a phase of liquid crystals that favours equally spaced layers and inherent molecular chirality. These two conditions cannot be simultaneously satisfied, resulting in frustration.

Frustration need not be an intrinsic property of the free energy functional.
In geometric frustration, the symmetries of the local groundstate or microstructure are not a subset of the symmetries of the manifold on which the system lives. In the smectic-A${^*}$ phase, molecular chirality is incompatible with equally spaced layers. A proliferation of phases attempt to mediate between these two extremes. Most notably, the twist grain boundary phase (TGB), employs grain boundaries created from an infinite row of parallel screw dislocations to rotate flat layers \cite{Renn:1988p2132}.

The helical nanofilament (HN) phase, like the TGB phase, originates from the intrinsic frustration between equally spaced layered smectic phase and macroscopic chirality. In systems of achiral bent-core liquid crystals and mixtures of achiral bent-core and rod-like molecules, chiral phases often arise from spontaneous symmetry breaking \cite{Link:1997p1924,Dozov:2001p247,Earl:2005p021706,Takanishi:2005p4020}. Macroscopic homochiral domains populate the sample, with both handednesses occurring with equal probability.  Unlike the B2 and B3 bent core phases, where molecular tilt with respect to the smectic layers admits chirality through the spontaneous breaking of mirror symmetry, the HN phase, a smectic A phase, allows the director to twist with respect to the layers \cite{Sekine:1997p1307,Sekine:1997p6455,Thisayukta:2001p3277,Niwano:2004p14889}.

Although initially considered to have the same morphology as the TGB, the hierarchical structure of the HN phase exhibits a distinctively different motif. Upon cooling from a high temperature fluid phase, helical bundles consisting of approximately five nested smectic layers, nucleate.  These homochiral, coherently rotating filaments form the basis of the hierarchical HN phase.  They assemble, with axes aligned, into a hexagonal lattice producing a nanoporous bulk structure. From freeze fracture experiments on the bulk HN texture, an archetypal Bouligand texture \cite{Bouligand:1984p1899,Livolant:1989p91} reveals an underlying cholesteric texture with the pitch direction parallel to the center of the filaments \cite{Hough:2009p456}.  The HN phase accommodates chirality by forming helicoidal layers that locally match a cholesteric texture at the expense of long ranged ordering of the layers.  Conversely, the TGB phase locally prefers flat layers, only admitting chirality across grain boundaries, thus allowing the layer normal to rotate by a fixed angle.

\section{The Smectic Free Energy}

Landau-de Gennes theory has proven exceedingly powerful in understanding the phase behavior of smectics, particularly the nematic to smectic-A transition (NA)  \cite{deGennes:1972p753}.  The onset of smectic order is characterised by the development of a mass density wave locally modulating the molecular positions.  In mean-field theory, a non-zero value for the complex smectic order parameter $\psi(\x)=\psi_0(\x) e^{i \qsm \phi(\x)}$ indicates both the emergence of smectic order and the location of smectic layers denoted by level sets of the phase field $\phi=n d$, where $d=2 \pi/\qsm$ is the layer spacing.  Gauge-like minimal coupling of the nematic director field $\n$ to the smectic order parameter, reminiscent of the Landau-Ginzburg theory for superconductors, penalises deviations of the director field from the layer normal $\N$. The low energy deformations to the nematic director field still cost energy in the smectic phase.  This phenomenology is captured by the Landau-de Gennes free energy,
\begin{eqnarray}
F_{\rm L-dG}&=&\int d^3x \Big\{C\big\vert(\nabla-i\qsm \n)\psi\big|^2+(t-t_c)\vert\psi\vert^2+\frac{u}{4}\vert\psi\vert^4\nonumber\\
&&+\frac{K_1}{2}\big(\nabla\cdot\n\big)^2+\frac{K_2}{2}\big(\n\cdot\nabla\times\n+\q\big)^2+\frac{K_3}{2}\big((\n\cdot\nabla)\n\big)^2 \Big\},
\end{eqnarray}
where $\q$ is the chirality of the high temperature cholesteric phase and the  three terms of the Frank free energy describe the splay, twist, and bend deformations of the nematic field. Further discussion shall be restricted to the London limit \cite{London:1935p71} (unless otherwise specified), where gradients in the magnitude of the smectic order may be neglected.

The ratio of two natural lengthscales, $\kappa_G=\lambda/\xi$, the twist Ginzburg parameter, governs the phase behavior of chiral smectics: the twist penetration depth is the maximum length the director can deviate from the layer normal, $\lambda=(K_2/B)^{1/2},$ where $B=2C\qsm^2\psi_0^2$ is the compression modulus, and the coherence length $\xi=(C/\vert t-t_c\vert)^{1/2}$ is the distance over which $\psi_0$ decays to zero.  Type I materials ($\kappa_G<\frac{1}{\sqrt{2}}$) completely expel chirality from the smectic phase.  In the smectic phase $\psi_0\ne0$, the free energy density attains its minimal value $f_{\rm sm_A}=\frac{\vert t-t_c\vert^2}{u}+\frac{K_2\q^2}{2}$ corresponding to the order parameter $\psi_0=(\frac{-2\vert t-t_c\vert}{u})^{1/2},$ and the cholesteric phase becomes favorable above the thermodynamic critical chirality $q_{\rm th}=(\frac{2}{K_2u})^{1/2}\vert t-t_c\vert.$

In the case of a mixed phase, containing both smectic and cholesteric qualities, a simple treatment no longer governs the thermodynamic properties. Deformations of the director field and layer spacing simultaneously contribute to the nature of the phase.  Restricting to regions within the smectic phase, the free energy reduces to
\begin{eqnarray}\label{L-dG energy}
F_{\rm L-dG}&=&\frac{1}{2}\int d^3x \left\{-\frac{(t-t_c)^2}{u}+B\vert\nabla \phi- \n\big|^2\right.\nonumber\\
&&\left.+K_1\big(\nabla\cdot\n\big)^2+K_2\big(\n\cdot\nabla\times\n+\q\big)^2+K_3\big((\n\cdot\nabla)\n\big)^2 \right\}.
\end{eqnarray}

\subsection{The TGB Phase}

The TGB phase, an intermediate phase featured in type II materials ($\kappa_G>\frac{1}{\sqrt{2}}$) admits chirality into the smectic via grain boundaries formed from an infinite row of parallel screw dislocations.  Each grain boundary joins two regions of smectic-A together by rotating the layer normals through an angle $\alpha=2 \tan^{-1}(a/\ell_d),$ which depends on the separation between defects $\ell_d$. A lattice of parallel grain boundaries separated by $\ell_b$ enable the smectic normals to approximately follow an underlying cholesteric texture with the average chirality $\bar{q}=\alpha/\ell_b.$  

Before calculating the free energy density of the TGB phase, the natures of both the layer morphology and director field need be made manifest.  The Euler-Lagrange equations for the free energy functional, Eq. (\ref{L-dG energy}), in the one elastic constant approximation $(K_1=K_2=K_3=K)$, given by
\begin{eqnarray}\label{var phi}
\frac{\delta F}{\delta \phi}&=&\nabla^2\phi-\nabla\cdot\n=0\\
\label{var n}\frac{\delta F}{\delta \n}&=&B(\nabla\phi-\n)+K(\nabla^2\n-2 \q\nabla\times\n)=0,
\end{eqnarray}
can be difficult to solve for generic director field $\n$.  If the director field is divergence free, the phase field will always be given by harmonic functions.  For instance, the phase field for a single screw dislocation, $\phi_{\rm screw}=z-\frac{b}{2\pi}\tan^{-1}(y/x),$ satisfies Laplace's equation everywhere except along the lower dimensional set $y=0, \, x=0$ which defines the dislocation core.

In the low chirality limit, only a single grain boundary need be considered \cite{Renn:1988p2132}, which is given by Scherk's first surface (shown in Fig. 1(a).) \cite{Kamien:2001p797}.  The $\pi/2$ TGB has the morphology of Schnerk's first surface \cite{Santangelo:2006p137801,Santangelo:2007p011702}.
We shall begin with the phase field for a single grain boundary
\begin{eqnarray}\label{grain_boundary}
\phi_0&=&\cos\left(\frac{\alpha}{2}\right)x-\frac{b}{2\pi} \Im \ln \cos\left(\frac{\pi}{\ell_d}(y+i z)\right)\nonumber\\
&=&-\frac{b}{2\pi}\Im \ln \left[e^{-\pi z /\ell_d}e^{-2\pi i\phi_- /b}+ e^{\pi z/\ell_d}e^{-2 \pi i \phi_+/b}\right],
\end{eqnarray}
that rotates layers at $z=-\infty,$ $\phi_-=x \cos(\alpha/2)-y \sin(\alpha/2)$ by $\alpha$ to $\phi_+=x \cos(\alpha/2)+y \sin(\alpha/2)$ at $z=+\infty,$ where $2 \sin(\alpha/2)=b/\ell_d,$ $-b$ is the winding of the phase around any of the line singularities, and $\ell_d$ is the spacing between defects within the grain boundary.  A branch point singularity marks each of the screw dislocations in the grain boundary at $z=0$ and $y=(j+1/2)\ell_d, \  j\in\mathbb{Z}.$  

\begin{figure}[h]
\centering
\includegraphics{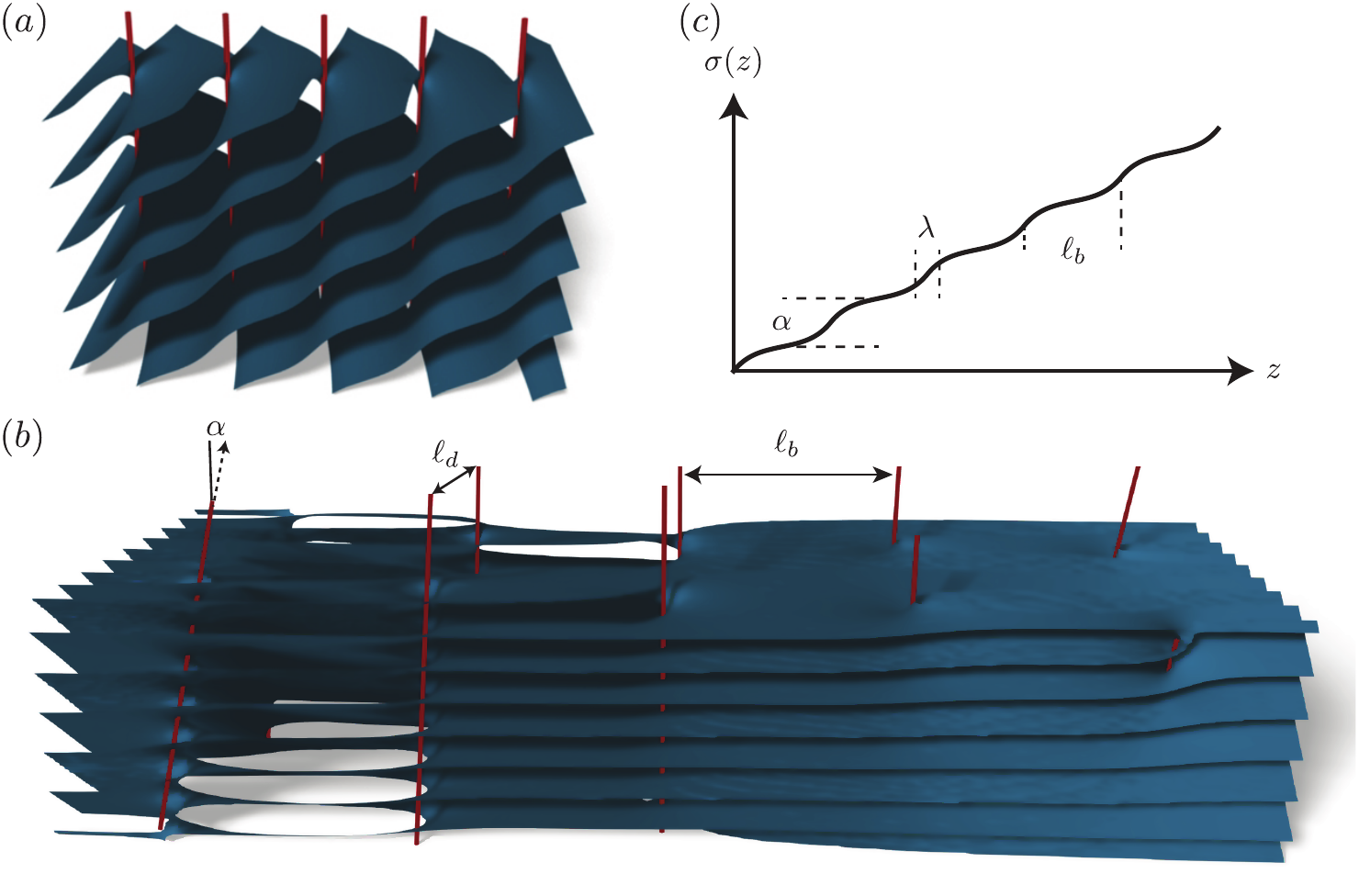}
\caption{(a) A single grain boundary consists of an infinite row of helicoidal dislocations along the $x-$axis, separated by spacing $\ell_d$. This rotates planar layers at $z=-\infty$ through an angle $\alpha$ at $z=+\infty$. (b) The bulk TGB phase results from combining an infinite row of these grain boundaries periodically along the $z-$axis, separated by $\ell_b$. Each grain boundary rotates the layer normals through angle $\alpha$, thus the grain boundaries, themselves must be rotated with respect to one another along the $z-$axis. (c)The director field $\n=\cos(\sigma(z))\ex+\sin(\sigma(z))\ey$ also rotates in accordance with the layers. It is close to constant within each grain and jumps through angle $\alpha$ across each grain boundary. The width of this jump is given by the penetration depth $\lambda$.}
\end{figure}

A TGB bulk phase consists of a periodic array of grain boundaries living in the plane perpendicular to the pitch direction $\bf{\hat{q}}$, each rotated by $\alpha$ about $\bf \hat{q}$ with respect to the previous grain boundary.  The corresponding set of singularities exist at $z=k \ell_b, \ y \cos(k\alpha)-x \sin(k\alpha)=(j+1/2)\ell_d, \ \forall \, j,k\in\mathbb{Z}.$  The natural extension of the phase field becomes 
\begin{equation}\label{TGB1}
\phi_{\rm TGB}=\frac{1}{2}\left(\phi_{+\infty}+\phi_{-\infty}\right)-\frac{b}{2\pi}\sum_{k=-\infty}^{\infty}\Im\ln \cos\left[\frac{\pi}{\ell_d}\left(y \cos(k\alpha)-x \sin(k\alpha)+i(z-k \ell_b)\right)\right],
\end{equation}
where $\phi_{\pm\infty}$ are the asymptotic values of the phase field as $z\rightarrow\pm\infty$.  
In place of specifying the asymptotic values $\phi_{\pm\infty}$ of the phase field, it often proves more convenient to specify the behavior at the origin, for instance taking that there is a grain boundary at $z=0$, with the axes of the screw dislocations parallel to the $x-$axis.
If we note that the asymptotic values may be written as $\phi_{\pm\infty} = \phi_{\pm 1} + \sum_{k=1}^{\infty}(\phi_{\pm(k+1)}-\phi_{\pm k})$, where $\phi_{\pm k} = x \cos((k-1/2)\alpha)\pm y\sin((k-1/2)\alpha)$, then the phase field for the bulk TGB with specified behaviour at the origin is given by
\begin{eqnarray}\label{TGB2}
\phi_{\rm TGB}&=&-\frac{b}{2\pi}\Im \ln \left[e^{-\pi z /\ell_d}e^{-2\pi i\phi_{-1} /b}+ e^{\pi z/\ell_d}e^{-2 \pi i \phi_{+1}/b}\right]\nonumber\\
&&-\frac{b}{2\pi}\sum_{k=1}^{\infty}\Im\ln \left[1+e^{2\pi (z-k\ell_b) /\ell_d}e^{-2\pi i(\phi_{k+1}-\phi_k) /b}\right]\nonumber\\
&&-\frac{b}{2\pi}\sum_{k'=1}^{\infty}\Im\ln \left[1+e^{-2\pi (z+k'\ell_b) /\ell_d}e^{-2\pi i(\phi_{-(k'+1)}-\phi_{-k'}) /b}\right].
\end{eqnarray}
The morphology of this structure is shown in Figure 1(b). This form of the phase field has the advantage as it highlights the exponential overlap between grain boundaries.

\subsection{The Energetics}

To simplify the following calculations, the compression term in the free energy may be split into two terms, $\vert\nabla \phi-\n\vert^2=(\vert\nabla\phi\vert^2-1)+2(1-\n\cdot\nabla\phi),$ where the first term enforces equal spacing of layers and the second may be viewed as an effective potential felt by the director field. The gradient of the phase field will be necessary for calculating the director field and the free energy, which, upon some manipulations, is given by
\begin{eqnarray}\label{grad phi}
\nabla\phi_{\rm TGB}=\sin\left(\frac{\alpha}{2}\right)\sum_{k=-\infty}^{\infty}\big[\left(-\sin(k\alpha)\ex+\cos(k\alpha)\ey\right){g_\perp}_k+{g_z}_k\ez\big]\\
\end{eqnarray}
where
\begin{eqnarray}
\begin{array}{rcl}{g_\perp}_k&=&\textstyle\frac{\sinh\left(\frac{2\pi}{\ell_d}(z-k\ell_b)\right)}{\cosh\left(\frac{2\pi}{\ell_d}(z-k\ell_b)\right)+\cos\left(\frac{2 \pi}{\ell_d}(y \cos(k\alpha)-x\sin(k\alpha))\right)}\vspace{1.5mm}\\
{g_z}_k&=&\textstyle\frac{\sin\left(\frac{2 \pi}{\ell_d}(y \cos(k\alpha)-x\sin(k\alpha))\right)}{\cosh\left(\frac{2\pi}{\ell_d}(z-k\ell_b)\right)+\cos\left(\frac{2 \pi}{\ell_d}(y \cos(k\alpha)-x\sin(k\alpha))\right)}.
\end{array}
\end{eqnarray}
First we compute the layer compression energy density
\begin{eqnarray}\label{fcomp}
\hspace{-5mm}\frac{F^{\rm comp}}{V}&=&\frac{B}{2V}\int d^3x\left(\vert\nabla\phi\vert^2-1\right)\nonumber\\
&=&\frac{B}{2V}\int d^3x\left[\sin^2\left({\textstyle\frac{\alpha}{2}}\right)\sum_{j,k=-\infty}^\infty\left(\cos\left((j-k)\alpha\right){g_\perp}_j{g_\perp}_k+{g_z}_j{g_z}_k\right)-1\right]\nonumber\\
&=&\frac{2B\sin^2\left(\frac{\alpha}{2}\right)}{\ell_b}\int_{z_-}^{\ell_b/2}dz\sum_{k=-\infty}^{\infty}\left[\coth\left(\frac{2\pi\vert z-k\ell_b\vert}{\ell_d}\right)-1\right]\nonumber\\
&=&B\sin^2\left(\frac{\alpha}{2}\right)\left\{\frac{\ell_d}{\pi\ell_b}\ln\left[\frac{\sinh\left(\frac{\pi\ell_b}{\ell_d}\right)}{\sinh\left(\frac{2\pi z_-}{\ell_d}\right)}\right]-1+\frac{2z_-}{\ell_b}\right.\nonumber\\
&+&\left.\sum_{k=1}^{\infty}\left(\frac{\ell_d}{\pi\ell_b}\ln\left[\frac{\sinh\left(\frac{2\pi(k\ell_b- z_-)}{\ell_d}\right)}{\sinh\left(\frac{\pi\ell_b(2k-1)}{\ell_d}\right)}\frac{\sinh\left(\frac{\pi\ell_b(2k+1)}{\ell_d}\right)}{\sinh\left(\frac{2\pi(k\ell_b+z_-)}{\ell_d}\right)}\right]-2+\frac{4z_-}{\ell_b}\right)\right\}\nonumber\\
&=&\frac{B\ell_d\sin^2\left(\frac{\alpha}{2}\right)}{\pi\ell_b}\left\{\ln\left[\frac{1-e^{-2\pi\ell_b/\ell_d}}{1-e^{-4\pi z_-/\ell_d}}\right]+\sum_{k=1}^\infty\ln\left[\frac{\left(1-e^{-4\pi(k\ell_b-z_-)/\ell_d}\right)\left(1-e^{-2\pi\ell_b(2k+1)/\ell_d}\right)}{\left(1-e^{-4\pi(k\ell_b+z_-)/\ell_d}\right)\left(1-e^{-2\pi\ell_b(2k-1)/\ell_d}\right)}\right]\right\}\nonumber\\
\end{eqnarray}
where the cutoff $z_-\sim\xi$ is the size of the dislocation cores.

A full treatment of this problem would employ a completely general director field $\n=\cos(\tau({\bf r}))(\cos(\sigma({\bf r}))\ex+\sin(\sigma({\bf r}))\ey)+\sin(\tau({\bf r}))\ez$, where $\sigma({\bf r})$ and $\tau({\bf r})$ and generic rotations of the director field about the ${\bf z}-$ and ${\bf x-}$axes, respectively.
However, we shall consider an effective model for the director field, $\n=\cos (\sigma(z))\ex+\sin(\sigma(z))\ey,$ as $\n\cdot\nabla\phi$ becomes integrable.
The resulting effective free energy for the director field is given by \footnote{Note that $\frac{1}{A}\int dx dy {g_\perp}_k=\sgn(z-k \ell_b).$ For $m\ell_b<z<(m+1)\ell_b,  \, m\in\mathbb{Z},$ the integrals of the $x$ and $y$ components of the gradient of the phase field simplify as follows, $\frac{1}{A}\int dx dy {\phi_{\rm TGB}}_x=-\sin(\alpha/2)(\sum_{k=-\infty}^{m}\sin(k \alpha)-\sum_{k=m+1}^\infty\sin(k\alpha))=2\sin(\alpha/2)\sum_{k=m+1}^\infty\sin(k\alpha)=\cos(\alpha/2)-2\sin(\alpha(m+1)/2)\sin(\alpha m/2)=\cos(\alpha(m+1/2))$, likewise $\frac{1}{A}\int dx dy{\phi_{\rm TGB}}_y=\cos(\alpha(m+1)/2)$.}
\begin{equation}\label{sine-Gordon}
F_{\rm eff}=\int dz\left\{\frac{K_2}{2}\left(\partial_z \sigma- \q\right)^2 +B \left(1-\cos(\sigma-\alpha_m)\right)\right\},
\end{equation}
where $\alpha_m=\alpha(m+1/2)$ is the piecewise constant grain angle of the $m^{\rm th}$ grain. As each of the grains are identical, after a shift and rotation, it is sufficient to consider $\sigma_0=\sigma-\alpha_0$ in the region $0\le z\le\ell_b,$ with boundary conditions $\sigma_0(\ell_b/2)=0$ and $\sigma_0(0)=-\alpha/2.$ The solution to this variant of the sine-Gordon equation is given by the elliptic amplitude function,
\begin{equation}
\sigma_0(z)=2 \,\am\left(\frac{z-\ell_b/2}{\sqrt{m}\lambda},-m\right)
\end{equation}
where the value of the elliptic modulus $m$ is set by the implicit equation $\am\left(\ell_b/(2\sqrt{m}\lambda),-m\right)=\alpha/4,$ shown in Fig. 1(c).  Free energy density for the director field is
\begin{eqnarray}
\frac{F^{\rm dir}}{V}&=&\frac{1}{\ell_b}\int_0^{\ell_b} dz \left\{ \frac{K_2}{2}\left(\frac{2\, \dn\left(\frac{z-\ell_b/2}{\lambda\sqrt{m}},m\right)}{\lambda\sqrt{m}}-\q\right)^2+2B\,\sn^2\left(\frac{z-\ell_b/2}{\lambda\sqrt{m}},m\right)\right\}\nonumber\\
&=&\frac{8B\lambda}{\sqrt{m}\ell_b}{\rm E}\left(\frac{\alpha}4,-m\right)+K_2 \q\left(\frac{\q}{2}-\frac{\alpha}{\ell_b}\right)-\frac{2 B}{m},
\end{eqnarray}
where the elliptic functions are defined as $\dn^2(z,m)=1-m\, \sn^2(z,m)$ with $\sn(z,m)=\sin(\am(z,m))$ and ${\rm E}(z,m)=\int_0^z dt\sqrt{1-m\sin^2(t)}$ is the incomplete elliptic integral of the second kind.

The transition between smectic A and TGB occurs at $q_{c_1}=\frac{\ln \kappa_G}{\sqrt{2} \kappa_G}q_{\rm th}$ when the energy gain from introducing twist outweighs cost of introducing a single defect. Calculating the boundary between the TGB and cholesteric phases requires an examination of the stability operator $M(\x,\x')=\displaystyle \frac{\delta^2 F({\bf x})}{\delta \psi^*({\bf x})\delta \psi(\bf{x'})}=2C\{-\nabla^2-\xi^{-2}+\qsm^2\vert\nabla\phi-\n\vert^2\}\delta(\x-\x')$. In the linear approximation, $\phi=\phi_{\pm j}$ within each grain, the eigenfunctions are those of a harmonic oscillator and the minimal eigenvalue is $\qsm q_0-\xi^{-2}$. Thus, the onset of order coincides with the first eigenvalue of the stability operator going negative at $q_{c_2}=\sqrt{2}\kappa_G q_{\rm th}.$ \cite{Renn:1988p2132}

\section{The High Chirality Limit and the HN Phase}

Conversely, the HN phase requires a high background chirality in order to form. Thus we consider quenching smectic layers from a background pure cholesteric phase given by $\n = \cos(q_0 z)\ex +\sin(q_0 z)\ey.$ Growing layers from this configuration tries to minimise the compression term of the free energy $F^{\rm comp}=\int d^3 x\vert \nabla \phi_{\rm HN}-\n\vert^2$. If the cholesteric director field is confined to the $xy-$plane, rotation in registry along the pitch direction will minimise compression along the $z-$axis. Under this assumption, the constrained Euler-Lagrange for this system are given by 
\begin{eqnarray}\label{euler}
\nabla_\perp\cdot(\psi_0\nabla\phi)=\nabla_\perp \cdot (\psi_0^2\n).
\end{eqnarray}
Being a pure twist configuration implies $\nabla \times \n = -q_0 \n$, which violates the integrability condition for a scalar field $\nabla \phi=\n$. Yet there can still be a lower dimensional subspace of points satisfied by this condition. The helicoidal field 
\begin{eqnarray}\label{singlebundle}
\phi^{\rm loc}  = x \cos(q_0 z)+y \sin(q_0 z) = r \cos(q_0 z-\phi),
\end{eqnarray} 
vanishes along a two-dimensional subsurface $y \cos(q_0 z)-x\sin(q_0 z)=0$,  absolute minimum of the Landau-de Gennes free energy. Therefore, a helicoidal bundle of radius $R$ has compression energy per unit length of $\pi B q_0^2R^4/8$, whilst the condensation energy of forming the smectic-A texture in the same volume is $\pi K_2q_0^2R^2/2$. Consequently, the natural size for a helicoidal bundle is $R_*=\sqrt{2}\lambda$, where $\lambda$ is the penetration depth for twist, or the length over which the layer normal can deviate from the director field. \cite{Matsumoto:2009p257804}

\subsection{The HN Phase Morphology}
In order to create a bulk phase, a two-dimensional lattice of bundles forms with all of the layers coherently rotating in registry  \cite{Hough:2009p456}. That requires the bundles condense from a background phase of uniform high chirality. We must now solve the Euler-Lagrange equations, Eq.~\ref{euler}, for a periodic arrangement. The resulting phase field, $\phi$ must be harmonic in both $x$ and $y$ and will generically be satisfied by  the \emph{ansatz} that 
\begin{eqnarray}\label{HNgeneric}
\phi=\Re [\Theta(w)e^{-iq_0z}],
\end{eqnarray}
where $\Theta(w)$ is an analytic function of the complex variable $w=x+iy$ \cite{Matsumoto:2009p257804}. The choice of $\Theta=w$ recovers the optimal local configuration, Eq.~\ref{singlebundle} and implies that all simple zeros in $\Theta(w)$ result in a helical bundle with the same handedness as the underlying chiral cholesteric field. However, for convenience, we introduce the dimensionless coordinate $\zeta=w/l$ to set the lattice constant for all cases to unity.

Analytic functions with a periodic array of zeroes are given by Jacobi elliptic functions, doubly periodic analogues of trigonometric functions (see Appendix A). Each elliptic function has a corresponding simple pole for each simple zero, (see Table~I). Combining the elliptic functions in different combinations and tuning the elliptic modulus can generate a function with zeroes lying at the lattice vectors of any 2D Bravais lattice. The HN phase with square lattice symmetry is given by Eq.~\ref{HNgeneric} with 
\begin{eqnarray}\label{HNsq}
\tilde{\Theta}^{\rm sq}(\zeta)=\frac{1}{k^\sq}\frac{\sn(k^\sq \zeta,m^\sq)}{\dn(k^\sq \zeta,m^\sq)},
\end{eqnarray}
where the normalization $k^\sq=2K(m^\sq)\approx1.854$ is twice the complete elliptic integral of the first kind and the elliptic modulus for the square lattice is $m^\sq=1/2.$ The zeroes sit on the square lattice given by $\zeta=\{r,s\}, \, \forall \{r,s\} \in \mathbb{Z}$, and the requisite singularities are a lattice of simple poles that live on the dual lattice $\zeta=\{r+1/2,s+1/2\}, \, \forall \{r,s\} \in \mathbb{Z}.$
Likewise, the hexagonal HN phase is given by
\begin{eqnarray}\label{HNhex}
\tilde{\Theta}^{\rm hex}(\zeta)=\frac{3^{1/4}}{k^\hex}\left(1+\sqrt{3}\frac{\cn^2(k^\hex \zeta,m^\hex)}{\dn^2(k^\hex \zeta,m^\hex)\sn^2(k^\hex \zeta,m^\hex)}\right)^{-1/2},
\end{eqnarray}
where the normalization $k^\hex=2K(m^\hex)\approx3.196$ and the elliptic modulus for the hexagonal lattice is $m^\hex=\frac{2-\sqrt{3}}{4}$. Level sets of the phase field generated for this lattice are shown in Fig. 2(a). The simple zeroes live on the triangular lattice  $\zeta=\{r+s/2,\sqrt{3}s/2\}, \, \forall \{r,s\} \in \mathbb{Z}$. For the hexagonal lattice, the complementary divergences are square root branch points that live on a honeycomb lattice   $\zeta=\{r+s/2,\sqrt{3}s/2+1/\sqrt{3}\}, \,\zeta=\{r+s/2+1/2,\sqrt{3}s/2+1/(2\sqrt{3})\}, \, \forall \{r,s\} \in \mathbb{Z}$.

\begin{figure}[h]
\centering
\includegraphics{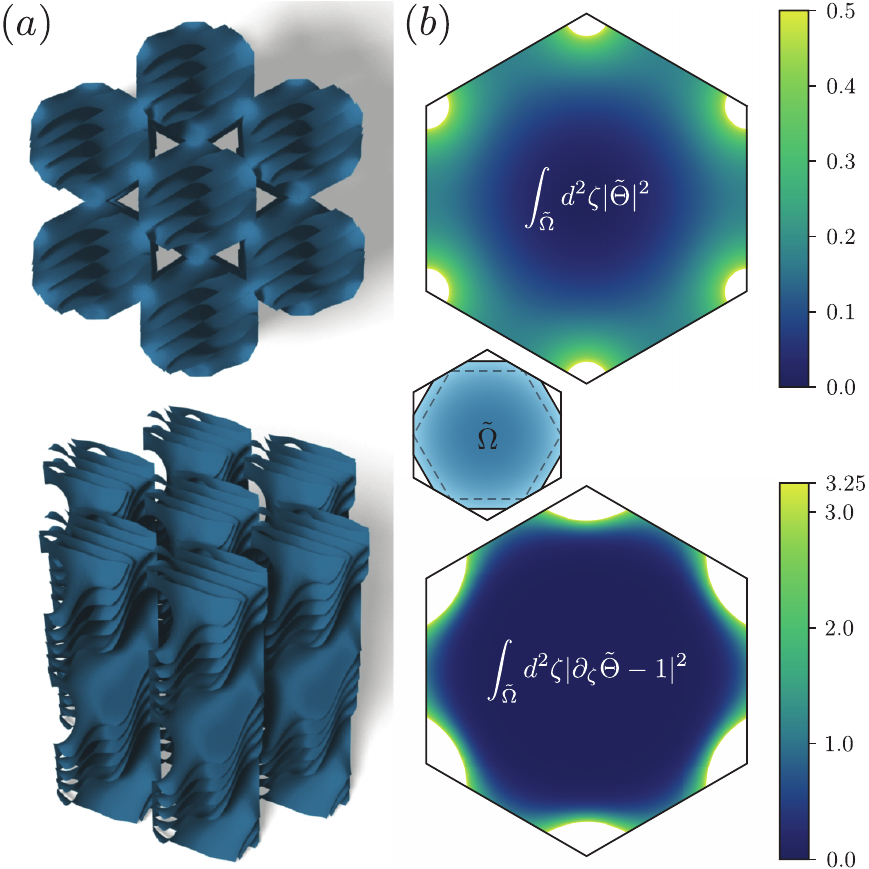}
\caption{(a) The HN phase consists of a nested array of coherently rotating helical bundles, close packed into a hexagonal lattice and is given by level sets of the phase field $\phi=Re[\tilde{\Theta}^{\rm hex}(\zeta)e^{iq_0z}]$. The centres of the bundles are simple zeroes in the function $\Theta^{\rm hex}(\zeta)$,  whilst branch point singularities at the corners of the fundamental hexagon diverge as $\zeta^{-1/2}$. (b) The compression energy for this phase field can be broken into two components $\int_{\tilde{\Omega}} d^2\zeta\vert\tilde{\Theta}^{\rm hex}(\zeta)\vert^2$ (top) and $\int_{\tilde{\Omega}}d^2\zeta\vert\partial_\zeta\tilde{\Theta}^{\rm hex}(\zeta)-1\vert^2$ (bottom). The inset shows the domain of integration, $\tilde\Omega$.
}
\end{figure}

\subsection{The Energetics}

The Landau-de Gennes free energy, Eq.~\ref{L-dG energy}, recast for the phase-field \emph{ansatz} $\phi=\Re [\Theta(w)e^{-iq_0z}]$ penalises deviations of the layer normal from the ideal cholesteric director field, $\n = \cos(q_0 z)\ex +\sin(q_0 z)\ey,$ in the compression term $F^{\rm comp}=B\vert\nabla\phi-\n\vert^2$. The compression energy density can be split into two components with $\nabla\phi \perp \n$ and $\nabla\parallel \perp \n,$ which, upon integrating out the director field, yield $f^{\rm comp,\perp}/(2\pi/q_0)=\frac{Bq_0^2l^2}{4}\vert \tilde{\Theta}\vert^2$ and $f^{\rm comp,\parallel}/(2\pi/q_0)=\frac{B}2\vert\partial_\zeta\tilde{\Theta}-1\vert^2$ respectively. The Landau-de Gennes free energy per volume is
\begin{eqnarray}\label{L-dG HN}
\frac{F}{V} = \frac{1}{A}\int_{\tilde{\Omega}}d^2\zeta\left\{\frac{Bq_0^2l^2}{4}\vert\tilde{\Theta}\vert^2+\frac{B}{2}\vert\partial_\zeta\tilde{\Theta}-1\vert^2-\frac{(t-t_c)^2}u\right\},
\end{eqnarray}
where $A$ is the total area of one unit cell of the lattice and $\tilde{\Omega}$ denotes the area occupied by the smectic phase.

By construction, both of the trial phases have a simple zero at the centre of each bundle, yet deviations from linearity with growing bundle radius combined with the divergences contribute to the compression energy, as these are the locales where the director field, corresponding to the layer normal deviates most from the ideal cholesteric. In the case of divergences, the layer normal traces out a helicoid of the opposite chirality and at the core of the divergence, the director would yield a helicoid of opposite handedness of the underlying cholesteric. These are points at which the smectic order breaks down, melting to the cholesteric. These melted regions have cholesteric energy proportional to their volume, giving the morphology of bulk phases the flavour of a packing problem.

The magnitude of the energies follows from the the higher order corrections to the Taylor expansions of $\tilde{\Theta}$ near centres of the bundles and the divergences. Near the simple zeroes, the expansions for the square and hexagonal lattices are given, respectively, by $\tilde{\Theta}^{\rm sq} = \zeta-\frac{{k^{\rm sq}}^4}{40}\zeta^5 +O(\zeta^9)$ and $\tilde{\Theta}^{\rm hex} = \zeta + \frac{{k^{\rm hex}}^6}{42\sqrt{3}}\zeta^7 + O(\zeta^{13})$, whilst in the vicinity of the poles, the Taylor series are, respectively, $\tilde{\Theta}^{\rm sq} = \frac{i}{k^{\rm sq}}\zeta^{-1}+\frac{{k^{\rm sq}}^2}{20}\zeta^3+O(\zeta^7)$ and $\tilde{\Theta}^{\rm hex} = \frac{(1-i)}{2}{p^{\rm hex}}^{3/2}\left(\zeta^{-1/2}-i\frac{3^{1/4}{k^{\rm hex}}^3}{6}\zeta^{5/2}\right)+O(\zeta^{11/2})$, where $p=3^{1/4}/k^{\rm hex} \approx 0.412$. It is clear the both terms will contribute to a higher compression free energy for the square lattice, thus we confine the remainder of our analysis to the hexagonal case. From the Taylor expansions near the zeroes and poles of the hexagonal lattice, the two components to the compression energy are
\begin{eqnarray}
&&\int_{\tilde{\Omega}}d^2\zeta\vert\tilde{\Theta}_{\rm hex}\vert^2 \approx 1-\frac{2 \times3^{1/4}}{k^{\rm hex}}\cosh^{-1}(2)\sqrt{1-\epsilon}\nonumber\\
&&\int_{\tilde{\Omega}}d^2\zeta\vert\partial_\zeta\tilde{\Theta}_{\rm hex}-1\vert^2 \approx \frac{3\sqrt{3}p^3}{\sqrt{1-\epsilon}}-4\sqrt{3}p+\epsilon+4p^{3/2}(12(1-\epsilon))^{1/4},
\end{eqnarray}
where $\epsilon$ is the filling fraction. (See Fig. 2(b).) Thus the total free energy for the hexagonal helical nanofilament phase is
\begin{eqnarray}
\frac{F}{V}=\frac{B q_0^2l^2}{\sqrt{3}}\left(1-\frac{2 \times3^{1/4}}{k^{\rm hex}}\cosh^{-1}(2)\sqrt{1-\epsilon}\right)+B\left(\frac{3\sqrt{3}p^3}{\sqrt{1-\epsilon}}-4\sqrt{3}p+\epsilon+4p^{3/2}(12(1-\epsilon))^{1/4}\right)-\epsilon\frac{(t-t_c)^2}{u}
\end{eqnarray}

\section{Discussion}
\subsection{The Phase Diagram}

Both the HN phase and the TGB phase condense from a higher temperature cholesteric, but the morphology of the HN phase is predicated upon having a higher chirality than is necessary to create the most highly chiral $\pi/2$ TGB phase. As the two constructions for the two phases follow from the same Landau theory, a phase diagram containing the two can be constructed. It behooves us to introduce a new non-dimensional parameter space, $\tilde{q}=q_0/\qsm$ and $\tilde{t}=(t-t_c)/(\qsm^2C)=1/(\xi^2\qsm^2),$ which yield free energy densities for the TGB and HN phases, respectively, given by
\begin{eqnarray}
&&\frac{F_{\rm TGB}}{BV}=\frac{\tilde{q}\sin\left(\frac{\alpha}{2}\right)}{\alpha}\ln\left[\frac{1-e^{-2\alpha\sin(\alpha/2)/\tilde{q}}}{1-e^{-4\sin(\alpha/2)/\sqrt{-\tilde{t}}}}\right]+\frac{8\tilde{q}\kappa_G}{\sqrt{m}\alpha\sqrt{-\tilde{t}}}E\left(\frac{\alpha}4,-m\right)+\frac{\tilde{q}^2\kappa_G^2}{2\tilde{t}}-\frac{2}{m}-\frac{\tilde{t}}{4}\left(1-\frac{2\tilde{q}}{\alpha\sqrt{-\tilde{t}}}\right)\\
&&\frac{F_{\rm HN}}{BV}=-\frac{\tilde{q}^2\kappa_G p^2}{4\tilde{t}}\left(1-2 p \cosh^{-1}(2)\sqrt{1-\epsilon}\right)+\frac{3\sqrt{3}p^3}{\sqrt{1-\epsilon}}+4p^{3/2}(12(1-\epsilon))^{1/4}-4\sqrt{3}p+\epsilon-\frac{\tilde{t}\epsilon}{4},
\end{eqnarray}
where $\kappa_G=\lambda/\xi$ is the Ginzburg parameter for the system, the TGB parameters are the turning angle, $\alpha$, and $m,$ the solution to $\am\left(\frac{\ell_b}{2\sqrt{m}\lambda},-m\right)=\alpha/4,$ and the HN parameters are the fraction of each unit cell occupied by smectic phase, $\epsilon$ and the hexagonal parameter, $p=3^{1/4}/(2 K(\frac{2-\sqrt{3}}{4}))\approx 0.412$. The border between the TGB phase and the cholesteric is given by the  upper critical chirality $\tilde{q}_{c_2}$, but the HN phase is stable for lower chiralities. The phase selection criterion compares the lowest of the TGB energies for all values of rotation angle $\alpha$ to the lowest HN energy for all possible filling fractions, $\epsilon$. The phase diagram is three dimensional with axes along the nondimensional chirality, $\tilde{q}$, reduced temperature $\tilde{t}$ and the Ginzburg parameter, $\kappa_G$. Slices with constant $\kappa_G$ are shown in Fig. 3(a). 

\begin{figure}[h]
\centering
\includegraphics{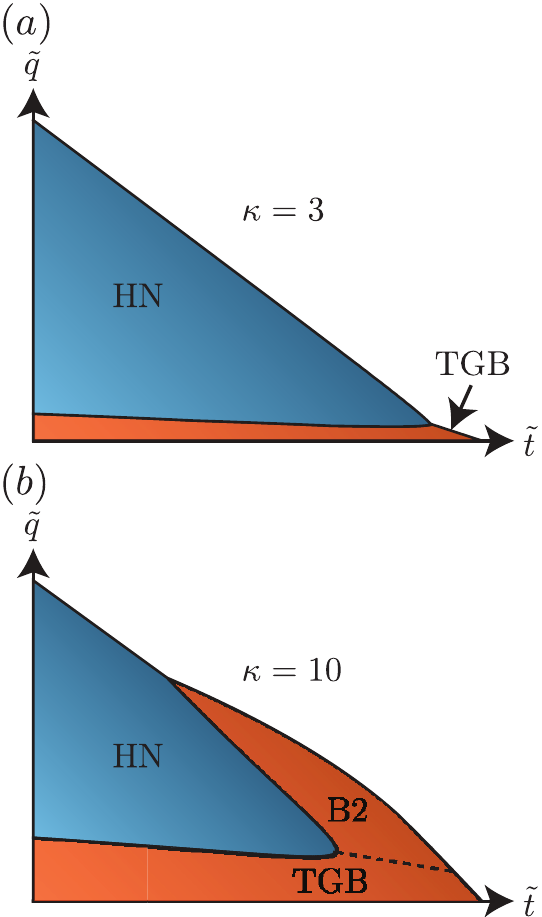}
\caption{Phase diagrams for the chiral smectic-A$^*$ phase depend on the reduced temperature, $\tilde{t}$, reduced chirality $\tilde{q}$ and Ginzburg parameter $\kappa_G=\lambda/\xi$ for the molecules. The phase boundary between the TGB and chiral nematic phases occurs at the upper critical chirality, $q_{c_2}$, whilst the onset of the HN phase occurs for lower reduced temperature and a finite, non-zero reduced chirality. (a) The phase diagram for a Ginzburg parameter $\kappa_G=3$. (b) For higher values of the Ginzburg parameter, the critical point for the onset of the HN phase moves deeper into the TGB phase. This leaves room for other high temperature bent core smectic phases, such as B2, from which HN experimentally condenses.}
\end{figure}

The most salient feature of this phase diagram is that  the HN phase only becomes stable above a critical chirality, which is smaller than $q_{c_2}$ and always occurs at lower reduced temperature than the zero chirality TGB/smectic/cholesteric triple point. As the Ginzburg parameter increases, the reduced temperature becomes lower and the TGB phase becomes reentrant upon increasing reduced chirality. (See Fig. 3(b).) Although this feature is unusual, it should be noted that in real systems the HN phase condenses from a higher temperature B2 phase of concentric cylindrical shells with a helical twist in the director field. This phase has not been included in our description and may account for the discrepancy.

\subsection{Conclusion and Future Directions}

We have presented a new continuum description for the smectic layer structure of the twist grain boundary phase, which features a continuously tunable angle $\theta$, the angle of rotation of the smectic layers across each grain boundary. This phase is one minimiser of the chiral Landau-de Gennes free energy. The helical nanofilament phase is another minimiser of the same free energy, stabile for higher chiralities than the TGB phase. There is one set of parameters for which the $\pi/2$ TGB phase and the square lattice description of the HN phase have identical smectic layers, although the underlying cholesteric reorients between the two phases. Our united description allows us to create a complete phase diagram for the Landau-de Gennes energy for chiral smectics-A. We demonstrate that the HN phase is generally stable for higher values of the reduced chirality and extends below the lower critical chirality of the TGB phase and into the cholesteric phase. This is analagous to the A$^*$ phase of helimagnets, which may be described by a similar Landau theory. Further, a complete description would include other bent core smectic phases, such as B2, and may shed light into exotic magnetic states.

\acknowledgements{This work was supported in part by NSF Grant DMR1262047 (RDK).  This work is also
partially supported by a Simons Investigator grant from the
Simons Foundation to R.D.K. G.P.A. was supported in part by the UK EPSRC through Grant No. EP/N007883/1.}

\end{document}